\documentstyle[epsfig,floats,tighten,preprint,prd,aps]{revtex}

\begin{document}

\preprint{CITA-2002-46, hep-th/0212327}

\title{Inflation and de Sitter Thermodynamics}

\author{Andrei Frolov and Lev Kofman}
\address{
Canadian Institute for Theoretical Astrophysics, 
University of Toronto\\
Toronto, ON, M5S 3H8, Canada
}

\date{\today}

\maketitle

\begin{abstract} 
We consider the quasi-de Sitter geometry of the inflationary universe.
We calculate the energy flux of the slowly rolling background scalar
field through the quasi-de Sitter apparent horizon and set it equal to
the change of the entropy (1/4 of the area) multiplied by the
temperature, $dE=T dS$. Remarkably, this thermodynamic law reproduces
the Friedmann equation for the rolling scalar field. The flux of the
slowly rolling field through the horizon of  the quasi-de Sitter
geometry is similar to the accretion  of a rolling scalar field onto a
black hole, which we also analyze.  Next we add inflaton fluctuations
which generate scalar metric perturbations. Metric perturbations result
in a variation of the area entropy. Again, the equation $dE=T dS$ with
fluctuations reproduces the linearized Einstein equations. In this 
picture as long as the Einstein equations hold, holography  does not put
 limits on the quantum field theory during inflation. Due to the
accumulating metric perturbations, the horizon area during inflation
randomly wiggles with dispersion increasing with time. We discuss this
in connection with the stochastic decsription of inflation. We also
address the issue of the instability of inflaton fluctuations
in the ``hot tin can'' picture of de Sitter horizon.
\end{abstract}

\pacs{PACS numbers: 04.50.+h; 98.80.Cq; 12.10.-g; 11.25.Mj}

\section{Introduction}\label{sec:intr}

The inflationary paradigm
 established during the last 20 years
assumes that the primordial equation of state is almost vacuum-like: 
$p\approx -\epsilon$.  To realize this equation of state, most models
deal with a scalar field $\phi(t)$  (or other fields which in
combination act as an effective scalar field) slowly rolling to the
minimum of its potential $V(\phi)$. During the slow roll regime the
homogeneous scalar field produces geometry which can be well
approximated by the quasi-de~Sitter metric. 

\begin{figure}[b]
  \centerline{\epsfig{file=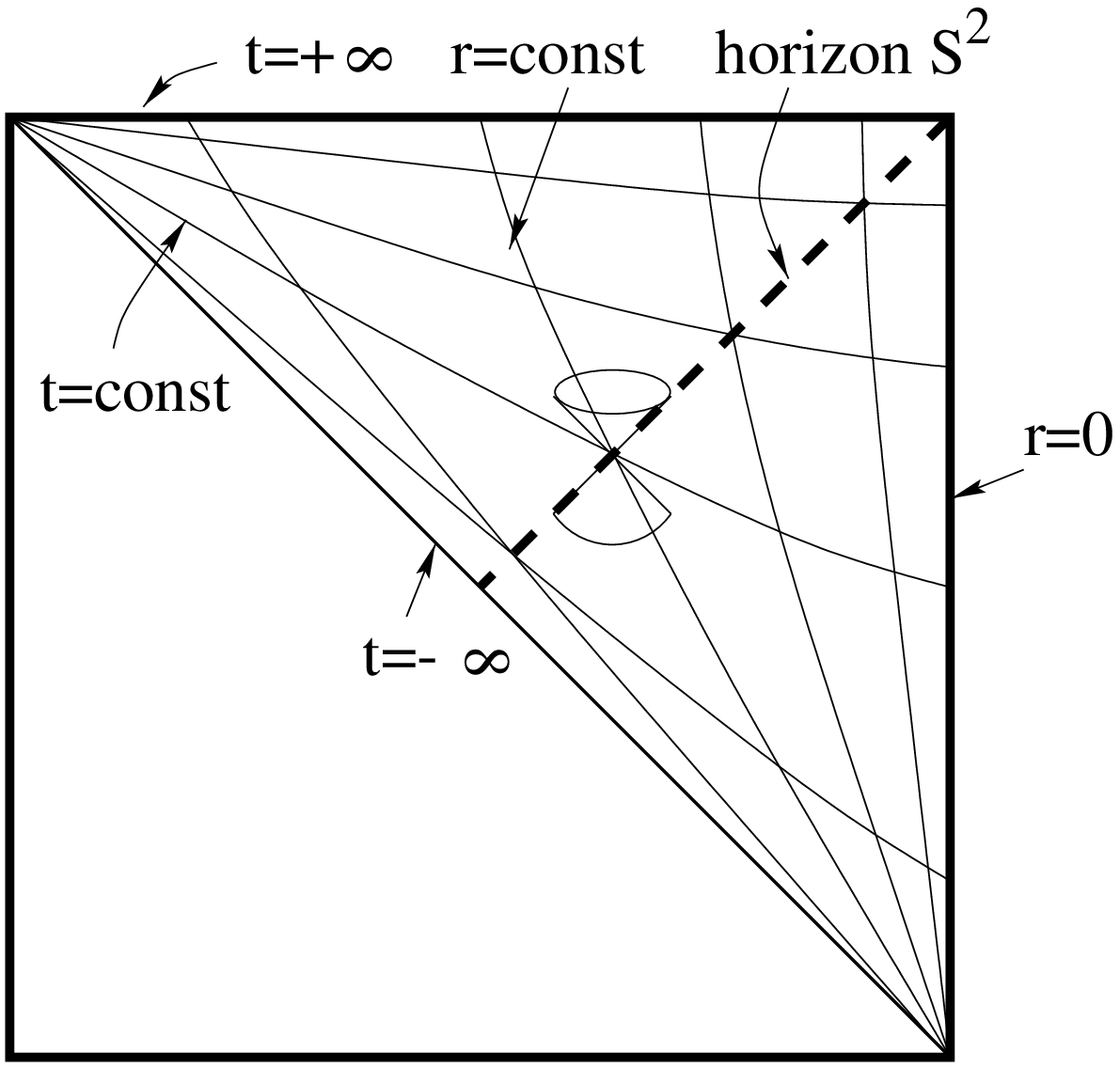, height=5cm}\hspace{2cm}
\epsfig{file=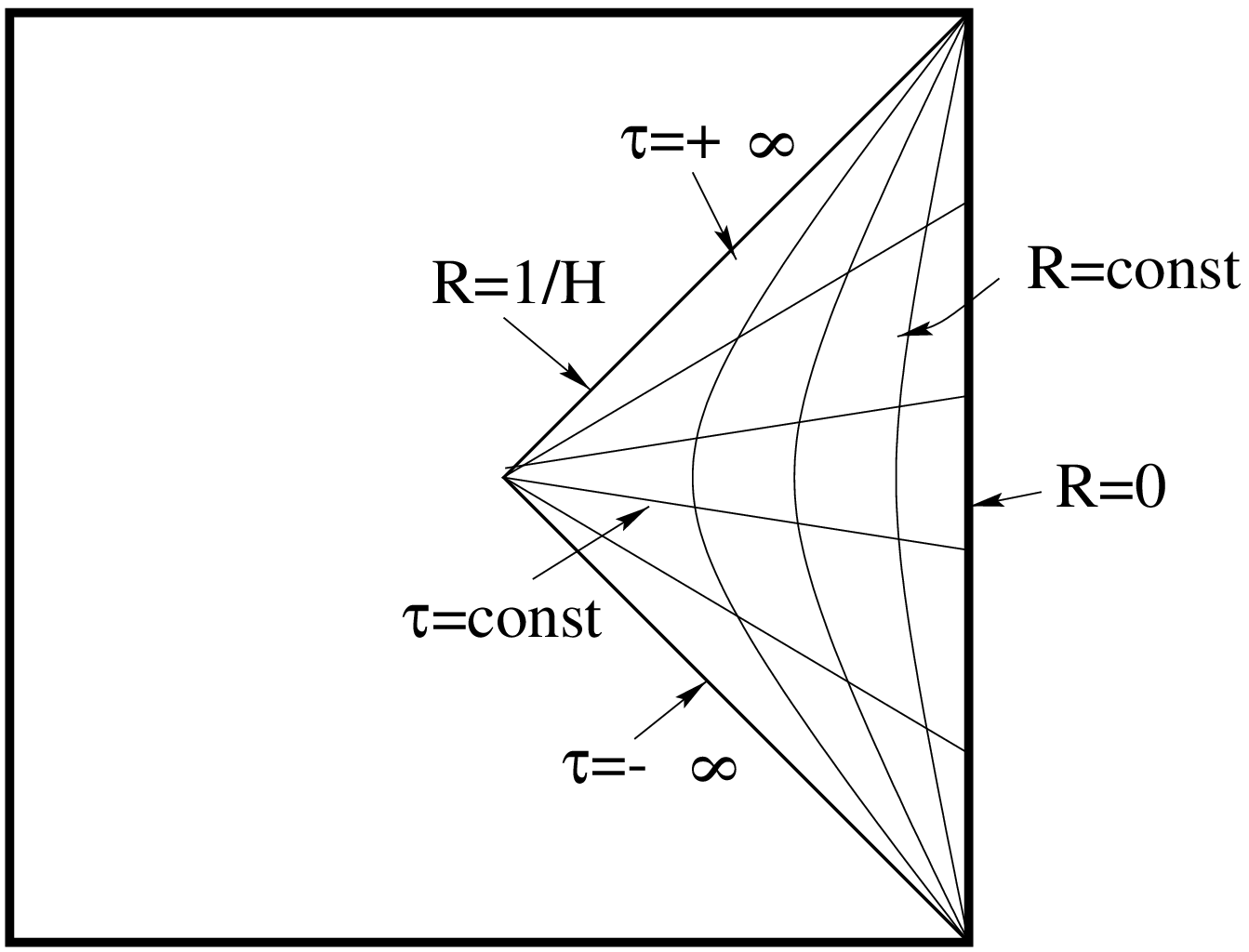, height=4.5cm}}
\medskip
  \caption{
    Penrose diagram of de~Sitter spacetime in the flat FRW coordinates
    (left) and the static coordinates (right). Each point represent a sphere
    $S^2$. Its radius at the horizon (dashed line on the left, edge of 
    diamond on the right) is equal to $\frac{1}{H}$.
}
\label{fig:penrose}
\end{figure}

The full pure de~Sitter spacetime, which corresponds to a 4d
hyperboloid of constant curvature,  can be compactly represented by its
Penrose diagram, given by the full square in Fig. \ref{fig:penrose}. It
can be covered by different coordinates. Cosmologists most often use
coordinates in which the metric is time-dependent  and corresponds to
an expanding flat universe
\begin{equation}\label{flat}
 ds^2= -dt^2 + e^{ 2Ht} \left(dr^2+r^2 d\Omega^2 \right) \ ,
\end{equation}
where $d\Omega^2=d \theta^2+\sin^2 \theta d \phi^2$. This coordinate
system covers the upper  half of the hyperboloid, which  corresponds to
the expansion branch.  The Penrose diagram of de Sitter spacetime in
flat FRW coordinates is shown on the left panel of Fig.~\ref{fig:penrose}.
Quasi-de~Sitter geometry is described by the scale factor $a(t)=a_0 \,
e^{\int dt H(t)}$, where the Hubble parameter $H$ is a slowly varying
function of time, $\dot H \ll H^2$.

The time-dependent form of the metric (\ref{flat}) is very convenient 
for investigating the dynamics of a scalar field with the equation
\begin{equation}\label{scalar}
\Box \phi=V_{,\phi}
\end{equation}
and for quantizing this field in the de~Sitter spacetime \cite{bd}.
Among quantum scalar fields with mass $m$ and conformal coupling $\xi$
in de~Sitter geometry, the case of minimal coupling $\xi=0$ and very
small mass $m \ll H$ plays an especially important role. Indeed, the
regularized vacuum expectation value is $\langle \delta \phi^2\rangle
=\frac{3H^4}{8\pi^2 m^2}$.  Formally, as was noted before the discovery of
inflation, this is an odd case since its eigen-spectrum contains an
infrared divergent term: $\langle \delta \phi^2\rangle \to \infty$ as
$m \to 0$. On the other hand, this is the most interesting case for
application to  inflation, since the theory of inflaton  (as well as
tensor) fluctuations is reduced exactly to this case.  Following the
time evolution of individual  fluctuations, it  was found  that the
infrared divergence can be interpreted  as  the  instability of quantum
fluctuations of  a very light  scalar field, which are accumulated 
with time $\langle \delta \phi^2 \rangle =\frac{H^3}{4\pi} \, t$
\cite{fluc,linde,star}. Fluctuations of $\delta \phi$ induce scalar
metric perturbations \cite{metr,star,hawking,guth,bardeen}.
This picture is a basis of the inflationary paradigm so successfully
confirmed observationally. Notice that heavy or conformal fields are
not produced by inflation. Further, backreaction of fluctuations
$\delta \phi$ leads to the picture of stochastic evolution of 
quasi-de~Sitter geometry \cite{stoch1,stoch2}, and at large values of
$H$ even to self-reproduction (eternal) of the inflationary universe
\cite{self}. Scalar field in the eternal inflationary universe is
described naturally in terms of the probability distribution function
$P(\phi, t)$ \cite{stoch2,stoch3}.

Recently, de~Sitter spacetime and inflation have drawn significant
attention in the theoretical physics/superstring community. Some of the
most interesting topics are holography and the thermodynamics
associated with the de~Sitter horizon.  In this context, the static
form of the metric of the de Sitter  spacetime
\begin{equation}\label{static}
ds^2=-(1-H^2 \, R^2)d\tau^2+
{{dR^2} \over{(1-H^2 \, R^2)}}+R^2 d\Omega^2 \ 
\end{equation}
 is commonly used.
The Penrose diagram of de Sitter spacetime in static coordinates is
plotted on the right  panel of Fig.~\ref{fig:penrose}. The classical
result of \cite{GH} is that observer at the origin detects a thermal
radiation from the de Sitter horizon at $R=\frac{1}{H}$ with the
temperature $T=\frac{H}{2\pi}$, and the horizon area
$A=\frac{4\pi}{H^2}$ is associated with the huge (geometrical) entropy
$S=\frac{A}{4G}$. Thermal vacuum in the causal patch (``hot tin can'')
corresponds to the Bunch-Davies vacuum of the metric (\ref{flat})
\cite{BMS,kleban} and gives a complementary picture of scalar field(s)
fluctuations. It is not clear to us, however, how quantum fluctuations
in the ``hot tin can'' picture correspond to the instability of 
quantum inflaton
fluctuations $\delta \phi$ and generation of metric perturbations. We
will return to this point at the end of the paper.

One of the issues in the holographic approach is the bookkeeping of
 entropy of
de~Sitter spacetime. The holography bound declares that the geometrical
entropy of the horizon exceeds the entropy of quantum states (of
 fields and particles)
within the volume surrounded by the horizon. It was recently claimed
that counting the entropy of quantum fluctuations generated during
inflation in the ``hot tin can'' and comparing it to the change of the
apparent horizon entropy violates the holography bound  unless an
ultra-violet cutoff of order of $\sim 10^{16}$ GeV in the momenta of
fluctuations is imposed \cite{kaloper}.

While it is expected that the approaches based on the time-dependent
form of the de Sitter metric with unstable fluctuations and the static
form of the de Sitter metric with thermal flux should give us
complementary insights, their languages are apparently different. This
is partly due to the difference between quasi-de~Sitter and  pure
de~Sitter geometries, and partly because different questions are
addressed. However, we have to understand how these two different
approaches to (quasi-)de~Sitter geometry with a scalar field   are
compatible with each other with respect to such important issues as the
generation of fluctuations,  entropy and global geometry.

In this paper we consider a particular question of how  the apparent
horizon area $A$, or the entropy $S=\frac{A}{4G}$, vary due to the slow
roll of the background scalar field and the generation of scalar metric
perturbations  during inflation. A novel element here is that we
combine the concepts of a dynamical, slowly rolling background field
and the instability of its fluctuations, with the concept of
geometrical, holographic  entropy.

In Section \ref{sec:geom}, we will calculate a variation of the 
geometrical  entropy  due to the energy flux through the apparent
horizon area. We find that, remarkably, the thermodynamical relation
$\delta E=TdS$ is equivalent to the Einstein equation for the rolling
inflaton field. In a sense, our  derivation of a  correspondence 
between thermodynamics and the Einstein equations for inflation is a
realization of such a correspondence 
found in an inspiring paper \cite{jacob} for local accelerating observers.
 However, we introduce a
technique to treat the apparent horizon of $R \times S^2$ topology
which is different from the description \cite{jacob} of a local Rindler
horizon for an accelerating observer.

As we will see, a non-vanishing flux is generated by the  kinetic term
$\dot \phi^2$ of the slowly rolling inflaton field.  It turns out that
this problem is very similar to the problem of the interaction of a
homogeneous rolling  scalar field  with a runaway potential $V(\phi)$
and a black hole.  In Section \ref{sec:bh}, we switch our attention
from inflation to black holes. A rolling scalar field  interacting with
a black hole is a  transparent illustration of the energy  flow of a
light scalar field through a horizon.

In Section \ref{sec:fluc1}, we return to inflation. On top of the
rolling background inflaton,  we consider  inflaton  fluctuations 
$\delta \phi$, which generate scalar metric perturbations $\Phi$. We
study the energy flux through the horizon including inhomogeneous
$\delta \phi$ fluctuations  and corresponding variations in the area of
the horizon, or entropy $dS$, which are sensitive to the scalar metric
perturbations $\Phi$. In this case, the calculations are more involved
than the calculations for the homogeneous time dependent background
field in  Section \ref{sec:geom}. This happens because there is no
exact Killing vector generating the horizon. However, for metric
perturbations which preserve  spherical symmetry we still can define
$TdS$ and compare it with the energy flux through the horizon.

In Section \ref{sec:fluc2}, we argue that the  metric perturbations
generated from inflaton quantum fluctuations can indeed be treated as
(locally)
spherically symmetric. We apply the general formalism for spherically
symmetric non-static geometry  of Section \ref{sec:fluc1} to
fluctuations from inflation. Again, we find that the thermodynamical
relation $\delta E=TdS$ leads to equations connecting $\Phi$ and
$\delta \phi$ which are in exact agreement with the linearized Einstein
equations for the fluctuations from inflation. This allows us to give new
insights into the entropy of cosmological fluctuations, as we will
discuss in Section \ref{sec:disc}.

\section{Slowly rolling inflaton and de Sitter entropy }\label{sec:geom}

We begin with an inflaton scalar field in the spacetime (\ref{flat}).
The homogeneous background field $\phi(t)$ is time-dependent and obeys
the equation (\ref{scalar}), which for the metric (\ref{flat}) reads
$\ddot \phi +3H \dot \phi +V_{,\phi}=0. $  If the Hubble parameter is
large, the friction term here is significant and  $\phi$ slowly rolls
towards the minimum of $V(\phi)$, realizing the inflation. 

The apparent horizon of de~Sitter spacetime has the topology of $R^1
\times S^2$, and is shown in Fig.~\ref{fig:penrose} in both coordinate
systems of interest. Imagine a spherical light front propagating from
outside towards  an observer at the origin $r=0$ or $R=0$. In the
time-dependent coordinates the space is exponentially expanding. There
is a critical, trapped surface $S^2$, which separates the region inside
the apparent horizon which is in causal contact with the observer, and
the outside region, where expansion keeps the light from reaching the
inside region.  The instant physical radius of an $S^2$ sphere  is
equal to $\rho=a(t)r$. Since the position of the apparent horizon is
$r=\frac{1}{a(t)H}$, its radius $\rho_{\text{h}}$ is equal to
$\frac{1}{H}$. It is constant in the pure de~Sitter space time, and a
slightly increasing function of time in the quasi-de~Sitter geometry
where $H(t)$ slowly decreases with time. 

So far we have not used the Einstein equations, which relate  the
energy momentum tensor  $T^{\mu}_{\nu}=\phi^{\mu}\phi_{\nu}-\left(
\frac{1}{2}\phi^{\sigma}\phi_{\sigma} -V(\phi)\right)\delta^\mu_\nu$ of
the background inflaton field $\phi(t)$ to the scale factor $a(t)$.
Actually, we are going to derive them  from the first law of
thermodynamics of the de~Sitter horizon.

Let us calculate the flux of energy through the area 
$A=\frac{4\pi}{H^2}$ of the apparent horizon of the quasi-de~Sitter
spacetime. In the spirit of the adiabatic approximation, we will
treat the horizon as a static surface for this purpose, but allow it to
slowly vary with time when calculating the change in its area. The
energy flux through the horizon is given by the integral
\begin{equation}\label{flux}
\delta E = \delta \int d\Sigma_{\mu} \, T^{\mu}_{\nu} \,  \xi^{\nu},
\end{equation}
where $d\Sigma_{\mu}$ is the 3-volume of the horizon, and $\xi^{\mu}$ is
the null generator of the horizon, as sketched in Fig.~\ref{fig:null}.
Notice that at the horizon $d\Sigma^{\mu}$ is parallel to $\xi^{\mu}$,
as both are null. We will equate the energy flux through the horizon to
the change of geometrical entropy $dS$
\begin{equation}\label{thermod}
TdS=\delta E= \delta \int d\Sigma_{\mu} \, T^{\mu}_{\nu} \,  \xi^{\nu} \ ,
\end{equation}
where the temperature $T$ is defined by the surface gravity at the horizon.
For the de~Sitter horizon, $T=\frac{H}{2\pi}$.

The entropy $dS$ and the expression (\ref{flux}) are geometrical
invariants and  can be calculated in arbitrary coordinates. In the case
of  de~Sitter geometry we can work either in time-dependent
(\ref{flat}) or static (\ref{static}) coordinates. The metric
(\ref{flat}) can easily be modified for the (adiabatic) regime of
slowly rolling $H(t)$, while it is not so straightforward for the
static form of the metric. Therefore, we will work in time-dependent
coordinates. Consider a slowly rolling inflaton $\phi(t)$ that is
time-dependent but homogeneous in the coordinates (\ref{flat}). The
background stress-energy tensor $T^{\mu}_{\nu}$ is diagonal. As the
(approximate) Killing vector is $\xi^{\mu}=(1, -H r, 0, 0)$, the energy
flux through the horizon $T_{\mu\nu} \, \xi^{\mu}\xi^{\nu}=\dot \phi^2$
is non-zero. To calculate the right hand side of equation
(\ref{thermod}), we integrate over the infinitesimal volume $4\pi \,
\frac{1}{H^2}\, d\lambda$, where $ \lambda$ is the affine parameter along
the null generators of the apparent horizon. In the left hand side of
(\ref{thermod}), the variation of the entropy is $dS = \frac{dA}{4G} =
-\frac{2\pi}{G} \, \frac{d H}{H^3}$. By comparing both sides of the
thermodynamic equation (\ref{thermod}) and dividing them by
$d\lambda=dt$, we obtain
\begin{equation}\label{einst}
\dot H=-4\pi G \, \dot \phi^2 \ .
\end{equation}
This is nothing else but one of the Einstein equations. Using the
equation of motion for $\phi$ and the equation (\ref{einst}), and
assuming the field $\phi$ is dominant, one can reconstruct the
Friedmann equation for $H$. Note that for a pure de~Sitter geometry
$\dot \phi=0$ and the energy flux through horizon is absent.

Formula (\ref{einst}) can also be derived from (\ref{thermod}) in the 
coordinates (\ref{static}). The coordinates $(t, r)$ of (\ref{flat})
and $(\tau, R)$ of (\ref{static}) are related by the transformations
\begin{equation}\label{transf}
r = e^{H\tau} \, \frac{R}{\sqrt{1-H^2R^2}} \ , \,\,\,\,\,\,
t = \tau+\frac{1}{2H} \ln (1-H^2R^2) \ . 
\end{equation}
In static coordinates, we have $\phi=\phi \left(\tau+\frac{1}{2H}
\ln (1-H^2R^2) \right)$. This $\phi(\tau, R)$ dependence corresponds to
the field profile delayed as $R$ approaches the horizon. We find a
similar delayed field configuration in black hole spacetimes in the
next Section. The energy flux through the horizon
$T^{\tau}_{R}=\phi_{,\tau}\phi_{,R} \sim \dot \phi^2$ is non-vanishing.
In these coordinates, however, a central observer does not see the flux
going through the horizon but rather accumulating the energy just
outside the horizon. To quantify this, one can define the local mass
function $M$ as it is done in Appendix B.

To conclude this section, we consider a sub-dominant, test scalar field
$\chi$ during inflation. We can calculate the flux of the field $\chi$
through the de~Sitter horizon. In this case, it is not important
whether we  are dealing with pure de~Sitter or quasi-de~Sitter
geometry.  Independent of the potential $V(\chi)$, the $\chi$-field
contribution to the energy flux $\delta E$ is given by precisely the
same expression as the inflaton field contribution. Either in terms of
the thermodynamic equation $TdS=\delta E$ or  in terms of the Einstein
equations (\ref{einst}), the contribution of $\chi$ is additive
\begin{equation}
\dot H=-4\pi G \, (\dot \phi^2 \ + \dot \chi^2) \ .
\end{equation}
This formula works even if the inflaton field velocity is sub-dominant,
$\dot \phi \ll \dot \chi$, and independent of the shape of the $\chi$-field
potential. In particular, it works for a massive scalar field $\chi$,
which oscillates about the minimum with an exponentially decreasing
amplitude.

\section{Rolling scalar field and Black Hole}\label{sec:bh}

In this section we interrupt the discussion of the slowly rolling
inflaton in the early universe and focus on the seemingly different
topic of a  scalar field accreting into a black hole. The reader of the
de Sitter story  can jump to the next Section.

Consider a homogeneous scalar field $\phi(t)$ with a runaway type
potential $V(\phi)$, say  $V \sim 1/\phi^2$ or $V(\phi) \sim
e^{-\phi/M_P}$. The concrete form of the potential is not important. We
can even broaden the class of models to include fields with the
potential $V(\phi)=\frac{1}{2}m^2 \phi^2$ with very small masses $m$,
as long as we deal only with the phase of  slow roll towards the
minimum. The well-known case of a massive scalar field  oscillating
around its minimum will have qualitatively different accretion onto a
black hole similar to the accretion of massive (quasi)-particles.

We will consider a test homogeneous (cosmological) scalar field
described by the equation (\ref{scalar}). In the vicinity of a black
hole, the spacetime is described by the Schwarzschild metric
\begin{equation}\label{static1}
ds^2=-\left(1-\frac{2M}{r} \right)dt^2+
{{dr^2} \over{(1-\frac{2M}{r})}}+r^2 d\Omega^2 \ ,
\end{equation}
and we will treat the cosmological evolution of the scalar field as the
boundary conditions imposed sufficiently far away from the black hole.
The scalar field retains its spherical symmetry $\phi=\phi(t, r)$,
so near a black hole it obeys the equation (\ref{scalar})
\begin{equation}\label{nearby}
  \left[\frac{\partial^2\ }{\partial r_*^2} - \frac{\partial^2\ }{\partial t^2}
    - \left(1-\frac{2M}{r}\right)\, \frac{2M}{r^3} \right] \Big(r \phi(r,t)\Big) -
 \left(1-\frac{2M}{r}\right)\, V_{,\phi }  
 = 0,
\end{equation}
where the tortoise coordinate $r_*$ is defined as $r_* = r+ 2M
\ln(r/M-2)$. The field asymptotics at the horizon $r \to 2M$
corresponds to an infalling wave $\phi(t+r_*)$. The boundary conditions
far away from the black hole follow from the assumption that there the 
cosmological field is spatially uniform and is free to roll down the
field potential. We take  $\phi(t, r)=\phi_{\infty} (t)$ at $r \gg
2M$, where the time dependent function $\phi_{\infty}(t)$ is the 
background evolution of the field in the absence of the black hole. 
The function $\phi_{\infty}(t)$ depends on the scalar field potential $V(\phi)$. 

Equation (\ref{nearby}) is well studied for the case of a massive scalar field
$V(\phi)=\frac{1}{2}m^2 \phi^2$, where it can be reduced to a linear
Schr\"odinger-type ordinary differential equation with an effective
potential $U_{\text{eff}}(r)$. 

\begin{figure}
  \centerline{\epsfig{file=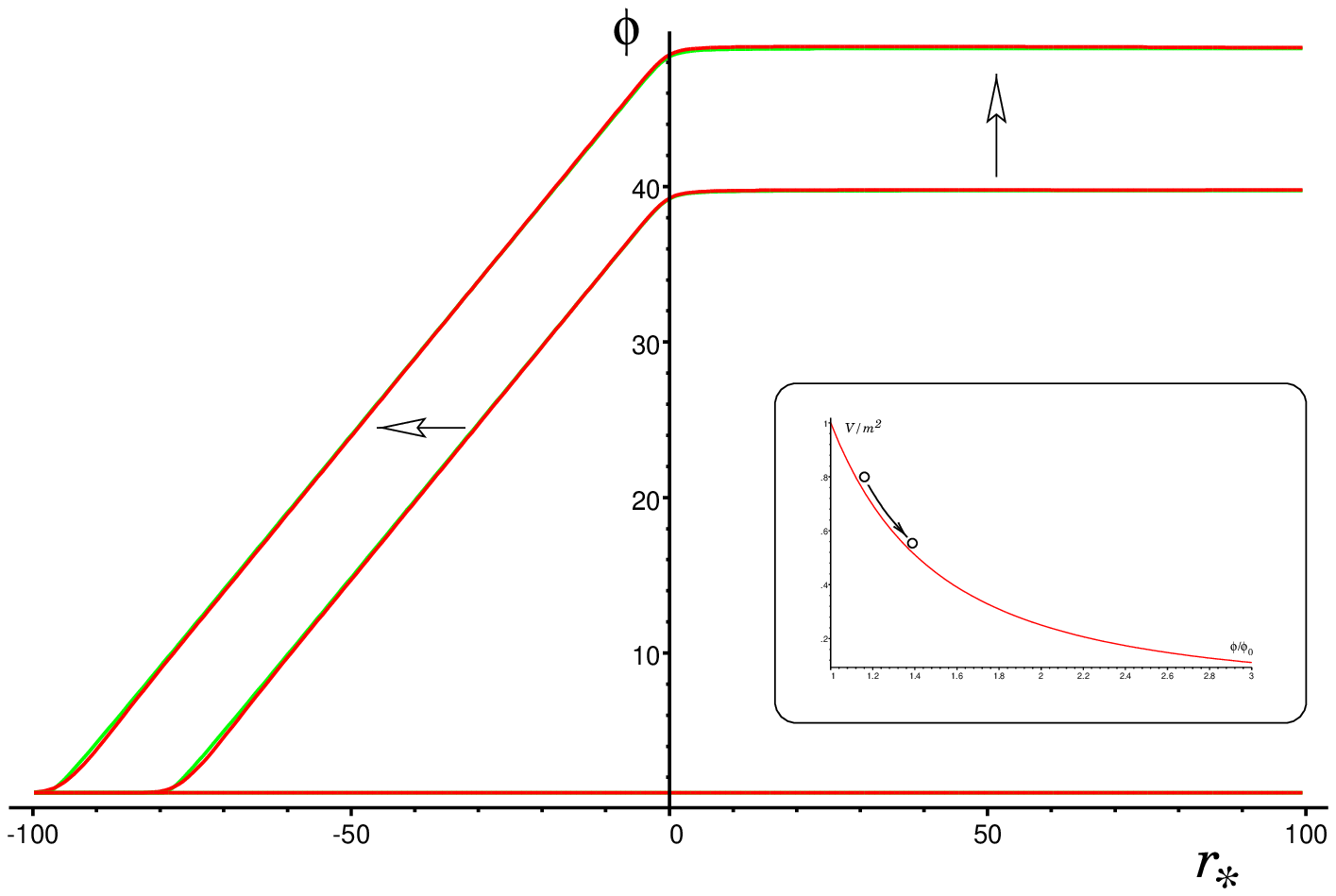, height=6cm}}
  \caption{
    Field profiles $\phi(t, r)$ outside a black hole for the runaway
    potential $V(\phi)=1/\phi^2$ at two subsequent time moments;
    horizontal axis is a tortoise coordinate  $r_*$. Red (solid) 
    curves are the numerical solutions, and green (pale) curves are
    obtained with the delayed field approximation. The horizontal line
    corresponds to the homogeneous initial conditions.
}
\label{fig:runaway}
\bigskip
  \centerline{\epsfig{file=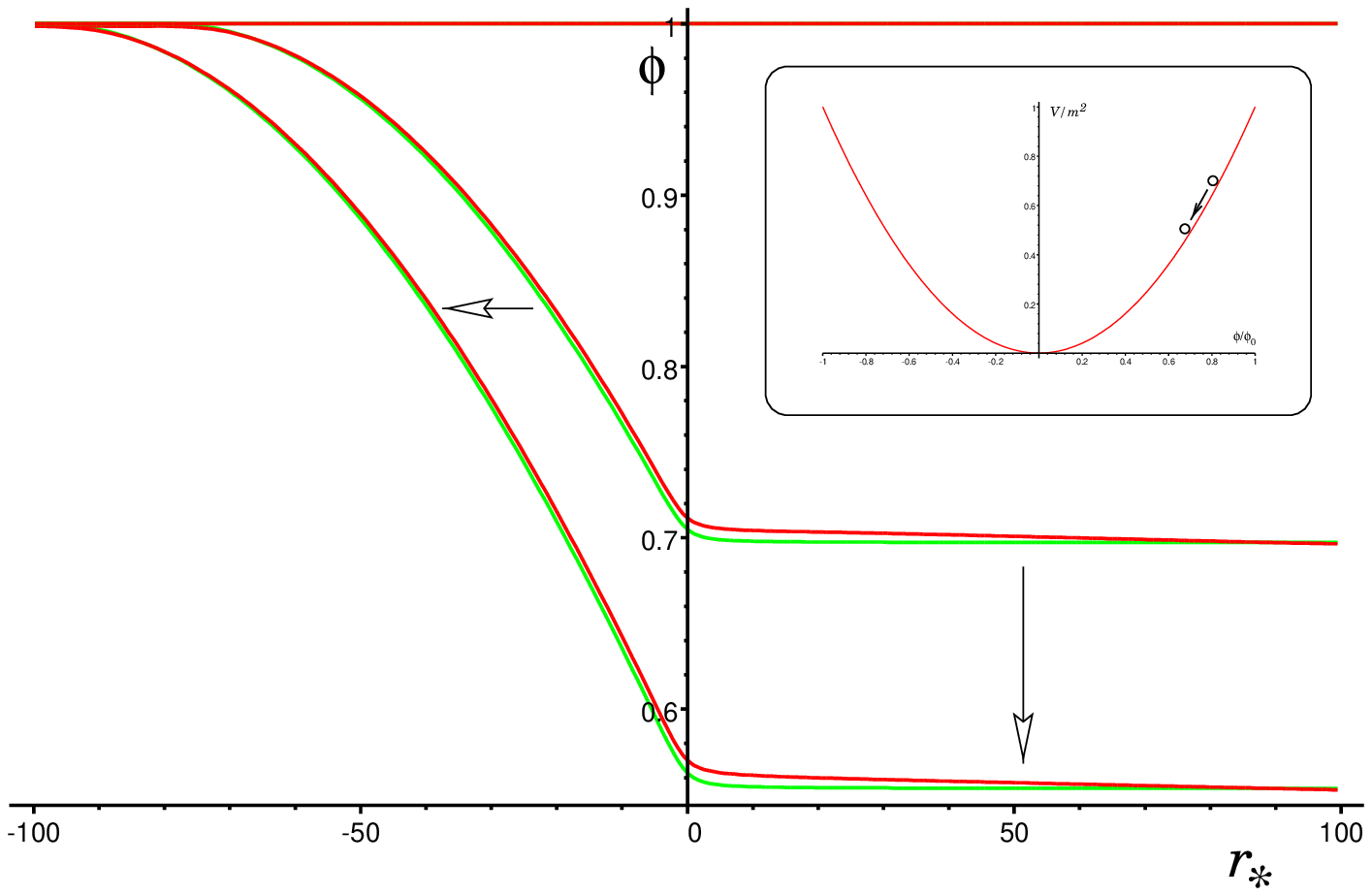, height=6cm}}
  \caption{
    Field profiles for the  potential $V(\phi)=\frac{1}{2}m^2 \phi^2$
    with a very small mass $m$ at two subsequent time moments. Red
    (solid)  curves are the numerical solutions, and green (pale)
    curves are the delayed field approximations.}
\label{fig:light}
\bigskip
  \centerline{\epsfig{file=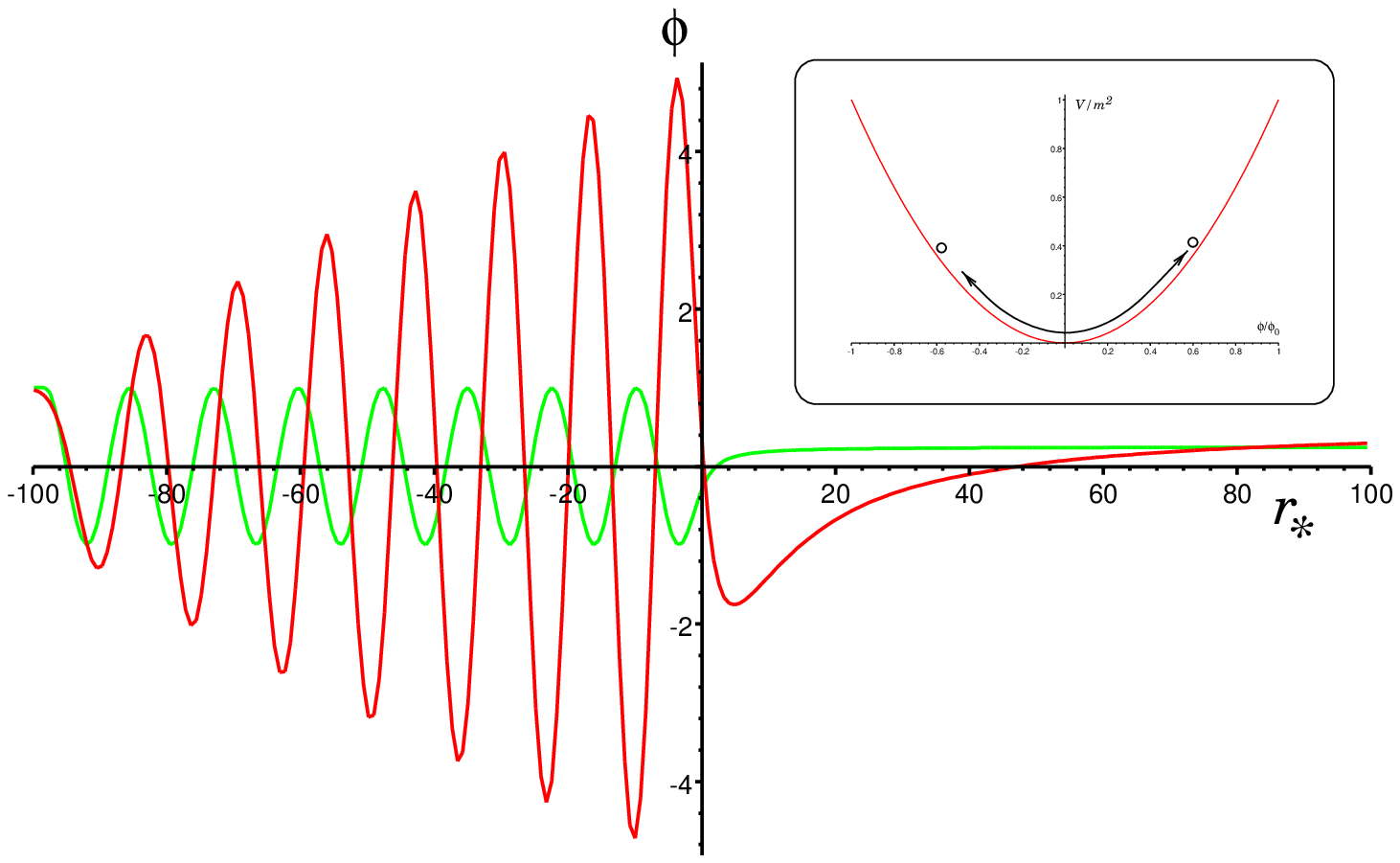, height=6cm}}
  \caption{
    A snapshot of the field profile for the  potential
    $V(\phi)=\frac{1}{2}m^2 \phi^2$ with a large mass $m$ when the
    field oscillates. Red (solid)  curve is the numerical solution, and
    green (pale) curve is the delayed field approximation.}
\label{fig:heavy}
\end{figure}

However, in general the equation (\ref{nearby}) is non-linear and the
behavior of its solution is very different from the case of a massive
scalar field; compare Figures \ref{fig:light} and \ref{fig:heavy}. We
treat equation (\ref{nearby}) as a partial differential equation and
solve it numerically for several examples of runaway potentials. We
also consider the case of a very light massive field and compare it
with the case of a heavy massive field. 

The sequence of radial profiles $\phi(t, r)$ for several moments $t$
is shown in Figure \ref{fig:runaway} for a runaway potential and in
Figure \ref{fig:light} for a very light massive field, respectively.

We found that $\phi(t, r)$ in these cases with a very high accuracy 
can be approximated by a simple formula
\begin{equation}\label{fit}
\phi(t, r)=\phi_{\infty} \left[
      t + 2M \ln \left(1 - \frac{2M}{r}\right)\right] \ .
\end{equation}
This result is rather insensitive to the form of the runaway potential.
We will call the solution (\ref{fit}) the delayed field approximation.
This is because the field profile is merely delayed, not frozen, on the
horizon. In the Eddington-Finkelstein coordinates $(v, r)$ the incoming
light geodesic is $v=t+r_*$. The solution (\ref{fit}) near the horizon
is $\phi \approx \phi(v - v_0)$ so the field crosses the horizon at the
speed of light, but is delayed by a small amount $v_0 = (2\ln 2 - 1)M$
compared to the null characteristic. Far away from the black hole, at
large $r \gg 2M$, a long-range tail is formed, which can be described
by the formula
\begin{equation}\label{tail}
\phi(t, r) \approx \phi_{\infty}(t)- \frac{(2M)^2}{r} \, \dot \phi_{\infty}(t) \ .
\end{equation}
For a light massive field when the equation (\ref{nearby}) is linear, 
the tail (\ref{tail}) can be related to scattering on the effective
potential $U_{\text{eff}}(r)$ with a sharp peak at $\approx 3M$.

Our approximation (\ref{fit}), verified by the numerical solutions,
generalizes the similar analytic result of \cite{jacob99} for the case
of a free scalar field ($V=0$), where $\phi_{\infty}(t)=t$. The
accuracy of the delayed field approximation can be estimated by
substituting the expression (\ref{fit}) into the differential equation
(\ref{nearby}) and calculating the residual term which is of order
${\cal O}\left( \frac{2M}{r}(1-\frac{(2M)^3}{r^3})\right)$. The residual
is bounded and localized at around $3M$; this is why the approximation
(\ref{fit}) works.

Equipped with equation (\ref{fit}), we can calculate the energy flux of
the scalar field through the horizon and the accretion rate onto the
black hole $\dot M=\frac{dM}{dt}$. We can calculate $\dot M$ in
different coordinates. In Schwarzschild coordinates, the spatial
gradient of $\phi$ develops just outside the horizon, and the energy
flux is defined by $T^t_r=\phi_{,t}\phi_{,r} \sim \dot
\phi_{\infty}^2$. In these coordinates, an observer at asymptotic
infinity does not actually see the energy flux go through the horizon.
What she/he sees is the energy from accretion accumulating just outside the
Schwarzschild radius, increasing the black hole mass (defined in the
ADM sense). In the Eddington-Finkelstein coordinates, the energy flux 
through the horizon is given by  $T_{vv}= \dot{\phi}_\infty^2$.

Independent of the choice of coordinates, the rate of accretion of the
scalar field $\phi$ by a black hole and the rate of increase of its
mass in the regime of the delayed field approximation is equal to
\begin{equation}\label{rate1}
  \dot{M} = 4\pi (2GM)^2 \, \dot{\phi}_\infty^2
\end{equation}
It is easy to check that this formula is equivalent to the first law of
black hole thermodynamics $\delta E=T dS$. We can rewrite (\ref{rate1})
in the form $ \left(\frac{1}{4GM}\right)^. =- 4\pi G \,
\dot{\phi}_\infty^2 $, which is readily similar to the corresponding formula
(\ref{einst}) for the slow roll scalar field in quasi-de~Sitter
geometry.

If we turn to the case of a massive scalar field with
$V(\phi)=\frac{1}{2}m^2\phi^2$, when the background scalar oscillates
around the minimum, we find that the delayed field approximation
(\ref{fit}) quickly fails, see Figure \ref{fig:heavy}, and formula
(\ref{rate1}) is not valid. In this case the scalar field can be
described in the WKB approximation. The accretion of the massive scalar
field looks like accretion of free massive quasi-particles of mass $m$.
Notice that  this is different from a massive oscillating scalar field
in the de~Sitter background, where the flux of the heavy  scalar field
through the horizon remains the same as for the light slow roll field,
$\sim \dot \phi^2$. 

In Appendix A, we briefly discuss some astrophysical applications of
the results of this section.

\section{Perturbed metric and horizon thermodynamics}\label{sec:fluc1} 

Classical perturbations in the de~Sitter geometry die out in accordance
with the ``no-hair'' theorem. Therefore, as in the black hole case,
there are  well-defined concepts of the entropy $S$ and temperature $T$
associated with the pure  de~Sitter horizon \cite{GH}.

The spherical symmetry of the spacetime naturally selects a foliation
by surfaces of the constant physical radius  $\rho=\text{const}$, with
normal vector $\nabla_{\mu}\rho$. If the spacetime is static (i.e
admits a timelike Killing vector field $k^\mu$ with vanishing
differential invariant $k_{\mu;\nu}k_\beta
\varepsilon^{\mu\nu\beta\alpha}$; do not mix with static form of the
metric), as de~Sitter spacetime is for example, the vector $\xi^{\mu} =
-\varepsilon^{\mu\nu} \nabla_{\nu}\rho$ orthogonal to
$\nabla_{\mu}\rho$ is also a Killing vector and a null generator of the
event horizon (which coincides with the apparent and Killing horizons
in the static case). Therefore one can define the surface gravity at
the horizon $\kappa$  by $\xi^{\nu} \xi_{\mu;\nu} = \kappa \xi_{\mu}$
and identify $\kappa=H$ and the temperature in the usual way
$T=\frac{H}{2\pi}$. The area entropy is equal to $1/4$ of the apparent
horizon area in the Planck units, $S=\frac{\pi}{G H^2}$.

For the  quasi-de~Sitter geometry with the homogeneous scalar field
$\phi(t)$ slowly rolling towards its minimum, when the horizon radius
$1/H(t)$ is adiabatically changing with time, we still can introduce an
entropy $S=\frac{\pi}{G} \, \frac{1}{H(t)^2}$ and temperature
$T=\frac{H(t)}{2\pi}$ associated with the apparent horizon. 

In inflationary cosmology, classical perturbations which could exist
prior to inflation die out. However, one of the   striking features of
the  quasi-de~Sitter inflationary  stage is that the quantum
fluctuations of the light (minimally coupled to gravity) scalar field
$\delta \phi$ are unstable and inevitably generate scalar metric
perturbations $\Phi(t, \vec x)$ \cite{metr,mukhanov}, so that the
quasi-de~Sitter metric (\ref{flat}) becomes perturbed
\begin{equation}\label{perturb}
 ds^2= -(1+2\Phi) dt^2+
 (1-2\Phi) e^{ 2 \int dt H} \left(dr^2+r^2 d\Omega^2 \right) \ .
\end{equation}

If the spacetime is not static (timelike Killing vector does not
exist), as in the inflationary geometry (\ref{perturb}), the situation
with the horizon thermodynamics is not quite so simple. The event
horizon and apparent horizon are in general different surfaces, and the
Killing horizon does not exist, so the notion of surface gravity is
ill-defined.

Fortunately, in the case of spherical symmetry one can define the mass
(energy) $M$ inside the spherical region of the physical radius
$\rho$ by $1-\frac{2GM}{\rho} = (\nabla\rho)^2$. The mass function $M$
obeys the mass formula \cite{pi} which follows from the Einstein equations
\begin{equation}\label{mass}
  M_{;a} = 4\pi\rho^2 (T_{ab} - T g_{ab}) \rho^{;b},
\end{equation}
where $a, b$ stands for $(t, r)$ only. We derive the mass formula 
(\ref{mass}) in Appendix B and give there other definitions to be used
in this Section.

We will argue now that the mass formula (\ref{mass}) applied at the
apparent horizon can be interpreted  as a first law of thermodynamics
$\delta E=TdS$ for an arbitrary spherically symmetric metric. In the
next section, we will apply this result to the perturbed inflationary
metric (\ref{perturb}), which as we will show can be well treated as
(locally) spherically symmetric. 

Let us look at the apparent horizon, as it is locally defined and much
easier to find than an event horizon. In spherical symmetry, the
position of the apparent horizon is given by $f \equiv (\nabla\rho)^2 = 0$.
The vector $\nabla_{a} f$ is normal to the apparent horizon, which is the surface
$f=0$, while the orthogonal vector $\zeta^a = \varepsilon^{ab} \nabla_b f$
is tangent to it. Unlike in the static case, these vectors are not
necessary null. The change of the mass function $M$ along the apparent
horizon is trivially related to the change of the radius of apparent
horizon,
\begin{equation}
  G \zeta^a M_{;a} = \kappa \rho \zeta^a\rho_{;a}
    = \frac{\kappa}{2\pi}\, \zeta^a (\pi\rho^2)_{;a} \ ,
\end{equation}
which allows its interpretation as the first law of thermodynamics
\begin{equation}\label{law}
\stackrel{\delta E}{\overbrace{\zeta^a M_{;a}}} \ = \ 
\stackrel{T}{\overbrace{\frac{\kappa}{2\pi}}} \ 
\stackrel{dS}{\overbrace{\zeta^a \left(\frac{ A}{4G}\right)_{;a}}} \ .
\end{equation}
The parameter $\kappa$ is defined in the Appendix B and generalizes the
surface gravity.  To relate the heat flow term with the stress-energy
tensor, we use the mass formula (\ref{mass}) which gives
\begin{equation}
  \zeta^a M_{;a} = 4\pi\rho^2 (T_a^b - T g_a^b) \zeta^a \rho_{;b}.
\end{equation}
The above expression can be rewritten in a more convenient form if one
realizes that $\zeta^a M_{;a} = - 2 \kappa \xi^a M_{;a}$
\begin{equation}\label{first}
    - \zeta^a \left(\frac{ A}{8\pi G}\right)_{;a}
    = 2{ A}\, T_a^b \xi^a \rho_{;b}.
\end{equation}
This formula relates the change of the apparent horizon area to the
(spherically symmetric) flux of matter through the $\rho=\text{const}$
surface passing through the point on the horizon at which $M$ is
evaluated.

\section{Fluctuations from inflation and wiggles of the horizon area}\label{sec:fluc2}

In the time-dependent metric (\ref{flat}) of the quasi-de~Sitter
geometry, quantum fluctuations of the inflaton scalar field can be
expanded with respect to the eigenmodes
\begin{equation}\label{QFT}
\hat \delta \phi(t, \vec x)= \sum_{lm} \int dk \,
 \left( \hat a_{klm} \,  \phi_{klm}+
\hat a_{klm}^+ \,  \phi_{klm}^* \right) \ ,
\end{equation}
where $\hat a_{klm}$ and $ \hat a^+_{klm}$ are annihilation and
creation operators; $\phi_{klm}=\phi_k(t)\, 2k j_{l}(kr) \, Y_{lm}(\vec
\Omega)$ are the eigenmodes. The $r-$dependent eigenmode factor is
given in terms of the  spherical Bessel functions $j_l(kr)$, $
Y_{lm}(\vec \Omega)$ are the spherical harmonics, and the
time-dependent eigenmode factor is given in terms of Hankel functions
$\phi_k(t)=\frac{\sqrt{\pi}}{2} \, H \eta^{3/2} \, {\cal H}^{(1)}(k
\eta)$, where $\eta$ is the conformal time $\eta=-\frac{1}{H}
e^{-Ht}$.  The Bunch-Davies vacuum corresponds to the absence of
particles in the past $\hat a_{klm} \vert 0 \rangle=0.$ This
corresponds to the positive-frequency asymptotic $\phi_k(t) \simeq
\frac{1}{\sqrt{2k}} e^{-ik \eta}$ in the far past $\eta \to -\infty$.
Quantum fluctuations of the inflaton field are unstable  and turn
into long-wavelength classical inhomogeneities. Indeed, the modes which
have physical wavelengths smaller than the horizon $1/H$, have $k \eta
\gg 1$ at the horizon and quickly oscillate without producing physical
effects. In contrast, the modes with the wavelengths which exceed the
horizon size  $k \eta=k \leq 1$ cease to oscillate, freeze out, and
look like a classical scalar field with amplitude $\phi_k(t) \approx
\frac{H}{\sqrt{2}k^{3/2}}$.

We need to know the form of the eigenmodes at the apparent horizon
where $r=\eta$. Again, the time-oscillating modes with small physical
wavelengths  $k r= k \eta \gg 1$ are spatially oscillating functions
$j_l(kr)$. Large  wavelengths modes with $k r= k\eta \ll 1$ have 
asymptotics $j_l(kr) \sim (kr)^l$. Notably, only the $s$-wave with
$l=0$ survives in this asymptotic. Therefore, in our discussion of the
wiggles of the horizon area due to metric fluctuations, we can consider
only spherically symmetric perturbations. 

Fluctuations $\delta \phi$ generate scalar metric perturbations  which
are represented in (\ref{perturb}) by scalar $\Phi(t, r)$. Now we are
going to relate $\delta \phi$ and $\Phi$ using  the apparent horizon 
thermodynamics.

In the perturbed  spacetime  (\ref{perturb}), the physical radius is
$\rho=(1-\Phi) \,  e^{\int dt H} \, r$. Apparent horizon is defined by
$(\nabla\rho)^2=0$.  From here we find for the  apparent horizon 
\begin{equation}\label{app}
  \rho_{\text{h}} = H^{-1} \left( 1 + \Phi + \frac{\dot{\Phi}}{H} - \frac{\Phi_{,r}}{\dot{a}} \right).
\end{equation}
The entropy of the apparent horizon and the temperature are identified
with geometrical quantities as
\begin{equation}
  S = \frac{ A}{4G} = \frac{\pi \rho_{\text{h}}^2}{G}, \hspace{2em}
  T = \frac{\kappa}{2\pi} = \frac{1}{4\pi \rho_{\text{h}}},
\end{equation}
where $\kappa$ is  defined by equation (\ref{temp}) of Appendix B.
Evaluating the equation (\ref{first}) relating the change in the
horizon area to the energy flux at the position of the apparent horizon
in the perturbed spacetime (\ref{perturb}), and keeping  only the
linear terms with respect to $\Phi$, $\delta \phi$, and their
derivatives, one finds the following expression for the perturbations
\begin{equation}\label{follow}
 \ddot{\Phi} - \frac{2}{a}\, \dot{\Phi}_{,r} +   \frac{1}{a^2}\, \Phi_{,rr}
 + H\dot{\Phi} =
    8\pi G\, \dot{\varphi} \left( \delta\dot{\varphi} -
 \frac{1}{a}\,   \delta\varphi_{,r} \right),
\end{equation}
valid at the position of the apparent horizon.

Even though we know already that the first law of thermodynamics 
(\ref{law}) originated from the Einstein equations in the form of
(\ref{mass}), and hence the above equation would follow from the
linearized  Einstein equations, let us explicitly demonstrate this
fact. The usual equations for metric  perturbations from scalar field
fluctuations are \cite{mukhanov} 
\begin{equation}
 \left(1 + \frac{\nabla^2_{\vec x}}{4\pi G\, a^2 \dot{\varphi}^2}\right) \Phi
=\left(\frac{\delta\varphi}{\dot{\varphi}}\right)^{.} \ ,
\hspace{2em}
  \frac{(a\Phi)^{\dot{}}}{a} = 4\pi G\, \dot{\varphi} \delta\varphi \ ,
\end{equation}
where the for spherically symmetric $\Phi$, $\nabla^2_{\vec x} \Phi =
\partial_{rr} \Phi + \frac{2}{r} \partial_{r} \Phi$. Combining both
equations by substituting them into an identity
$  2 \dot{\varphi} \delta\dot{\varphi} =
    \dot{\varphi}^2 \left(\frac{\delta\varphi}{\dot{\varphi}}\right)^{.}
    + (\dot{\varphi} \delta\varphi)^{\dot{}},$
 we find that
\begin{equation}
\ddot{\Phi} - \frac{2}{a}\, \dot{\Phi}_{,r} +   \frac{1}{a^2}\, \Phi_{,rr}
 + H\dot{\Phi} + \frac{2}{a^2} \left(\frac{1}{r} - \dot{a}\right) \Phi_{,r}=
 8\pi G\, \dot{\varphi} \left( \delta\dot{\varphi} -
 \frac{1}{a}\, \delta\varphi_{,r} \right) \ .
\end{equation}
This equation holds everywhere in the perturbed spacetime. When
evaluated at the apparent horizon, the last term in the left-hand side
vanishes and we are left exactly with the equation (\ref{follow}).

\begin{figure}
\centerline{\epsfig{file=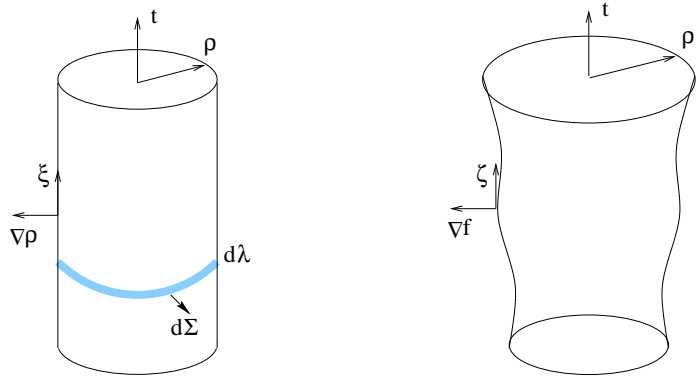, height=5cm}}
\medskip
\caption{
  A sketch of the apparent horizon of radius $\rho$. Left: the horizon
  in quasi-de~Sitter geometry where $\rho=\frac{1}{H}$ very slowly
  increases with time. Right: the same but with small metric
  perturbations where $\rho=\frac{1}{H} (1+\Phi)$ is wiggling with time
  due to the randomly varying $\Phi$. }
\label{fig:wigg}
\label{fig:null}
\end{figure}

Consider the area of the apparent horizon $A=4\pi \rho_h^2$. For frozen
long-wavelength fluctuations, we can drop derivatives in formula
(\ref{app}) for $\rho_{\text{h}}$. Then the leading contribution of 
the  frozen classical fluctuations to area is
\begin{equation}\label{wiggles}
A=\frac{4 \pi}{H^2} \,
\left(1+2\Phi \right) \ .
\end{equation}
Thus, the quantum fluctuations of the inflaton field  generate wiggles
of the quasi-de~Sitter horizon area, as described by (\ref{wiggles})
and sketched in Figure \ref{fig:wigg}.

Locally (within the  Hubble patch), almost spherical metric
perturbation $\Phi$ can be viewed as the variation of the Hubble
parameter $H$ \cite{star,stoch3}. Indeed, transform the metric
(\ref{perturb}) to synchronous coordinates $\tilde t=t+\int^t dt \Phi$,
$\tilde x_i=x_i+\int^t \frac{dt}{a} \int \nabla_i \Phi$. Then we have
$ds^2=-d \tilde t^2+e^{2\int d\tilde t \tilde H} d \tilde x^2$, where
the new Hubble parameter is  $\tilde H=(1-\Phi) H$. As expected, the
area of the horizon  is the same 
\begin{equation}\label{wiggles2}
A=\frac{4 \pi}{H^2} \,
\left(1+2\Phi \right) = \frac{4 \pi }{\tilde H^2}   \ .
\end{equation}
Therefore an observer within a Hubble patch sees the local
Hubble parameter $\tilde H$ simultaneously slow rolling
and wiggling due to the instability of quantum fluctuations
of the inflaton field, in the spirit of stochastic picture of inflation.

We have to note that there is another, much smaller correction to the
geometrical entropy due to the back reaction of quantum field effects
in curved background. Indeed, quantum gravity effects give 
one-loop corrections of order of $R^2 \sim H^4$ in the right hand side
of the Einstein equations, which slightly shift the Hubble parameter
\cite{KLS85} and the horizon area by a factor $1+c \frac{H^2}{M_p^2}$,
where $c$ is some numerical coefficient. However, this leads to a tiny
correction in the area entropy. 

Let us discuss the formula (\ref{wiggles}). 
Metric fluctuations 
$\Phi(t, r) \sim \delta \phi$ have a flat spectrum $\Phi_k=C \,
\frac{V^{3/2}}{M_p^3 V_{,\phi}} \, \frac {1}{k^{3/2}}$, where 
$C=\frac{16}{15} (3\pi)^{3/2}  $ is a
numerical coefficient.  Classical fluctuations of the inflaton field
$\delta \phi$  can be treated as a random (Wiener)  process with zero
vacuum expectation value, but with the increasing dispersion
$\langle\delta \phi^2\rangle \sim \int^{a H} \frac{dk}{k}=
\frac{H^3}{4\pi^2} t$.
Suppose for a moment that the 
background value of $H(t)$ is changing very slowly. Then,
  scalar metric fluctuations $\Phi(t,r)$ have zero mean value but
 their dispersion is increasing with time 
linearly, $\langle \Phi^2\rangle = C^2 \, \frac{V^3}{M_p^6 V_{,\phi}^2}
\int^{a H} \frac{dk}{k} \sim t$. Thus, the area of the apparent
horizon  looks like a random (Wiener) process with the mean value $\bar
A=\frac{4\pi}{H(t)^2}$ and with  the dispersion $\langle \Phi^2
\rangle$  linearly increasing with time. Imagine an ensemble of
inflationary Hubble patches with apparent horizon areas given by the
formula (\ref{wiggles}). Then their area is a statistical value which
obeys the Gaussian distribution $P(A, t)\sim e^{-\frac{(A-\bar
A)^2}{2\langle\Phi^2\rangle}}$. Now, if we take into account
 variation of background values like $H(t)$
and $V(\phi(t))$ with time, time-dependence and statistical
properties of the horiazon area given
by (\ref{wiggles}) or equivalently (\ref{wiggles2})
will be more complicated. We will discuss this issue below.

\section{Discussion: Entropy and Cosmological Fluctuations}\label{sec:disc}

In this paper we calculated the apparent horizon area and geometrical
entropy of quasi-de~Sitter geometry, which describes an inflationary
stage, and the energy flux of the scalar fields  through the apparent
horizon.  Issues related to de Sitter thermodynamics  are commonly
considered in the  static de Sitter coordinates, where an observer at
the origin is surrounded by the event horizon and detects a thermal
flux of temperature $T=\frac{H}{2\pi}$.

We work with the geometrical entropy $S$  of the apparent horizon  in
the time-dependent planar de-Sitter coordinates adopted in inflationary
cosmology, which admit a simple generalization to the quasi-de~Sitter 
geometry and, most importantly, admit a clear interpretation of
fluctuations generated from inflation.

The slow roll of the inflaton field leads to slow change in the horizon
radius. Assuming that the energy flux of the rolling scalar field
through the horizon  changes the geometrical entropy $\delta E=TdS$, we
reproduce the Einstein equation (\ref{einst})  which relates $\dot H$
and $\dot \phi^2$. Other background scalars $\chi$ which are
subdominant during inflation give similar contributions to the energy
flux $\dot \chi^2$, independent of their potentials.

This change of the quasi-de~Sitter horizon radius due to the rolling
scalar field is very similar to the increase of the mass of a black
hole due to the accretion of a background scalar field rolling  towards
the minimum of its potential $\dot M \sim \dot \phi^2$. This type of
accretion, which can be described analytically with the delayed field
approximation, can be realized for runaway potentials of the background
scalar field or for very light massive fields. We present this material
here mainly to illustrate similar calculations  for rolling scalars in
quasi-de~Sitter geometry, as the astrophysical effects of accretion of
a rolling scalar onto a black hole that we considered (say quintessence
and astrophysical-size black holes) are negligibly small. Oscillating
heavy scalar field accretes onto black hole very differently as
quasi-particles.

Inflationary, quasi-de~Sitter stage erases classical inhomogeneities
which could exist prior to it. However, quantum fluctuations of the
inflaton field (as well as other light scalars minimally coupled to
gravity) are unstable and produce long-wavelength fluctuations of the
inflaton field $\delta \phi$ which behave as classical inhomogeneities
at scales larger than the Hubble patch of size $H^{-1}$. Fluctuations
$\delta \phi$ generate long-wavelength scalar  metric perturbations
$\Phi$. Geometrical quantities, like an area of the apparent horizon,
acquire corrections due to the scalar metric perturbations. We
calculate the energy flux of the inhomogeneous scalar  field
$\phi(t)+\delta \phi(t, \vec x)$ through the apparent horizon and the
change in the apparent horizon area of the perturbed metric multiplied
by its (geometrical) temperature $T dS$. Again, equating $\delta
E=TdS$, we show that this thermodynamics relation is compatible with
the linearized Einstein equations which relate $\Phi$ and $\delta
\phi$.

Thus, as long as the Einstein equations hold, generation of the
inflaton fluctuations is in full agreement with the variation of the
entropy of the quasi-de~Sitter horizon. Therefore,
in the picture of rolling inflaton with quantum fluctuations generated 
 with time, we did not find that quantum fluctuations
may violate holographic bound  during inflation.
(Notice that the fluctuations from inflation
are described by squeezed states, which do not carry entropy
\cite{bpm,Kiefer}. In simple terms, locally the effect of fluctuations 
is just a  wiggling of the local Hubble parameter).

It was suggested in \cite{kaloper} that in the ``hot tin can'' picture 
 the entropy of fluctuations may violate holographic
bound  during inflation.\footnote{If frozen fluctuations from
 inflation were carrying large
entropy, then    a number $N$ of free light scalars
produced during inflation would have $N$ times bigger entropy and UV
cutoff proposed in \cite{kaloper} would depend on $N$.}
 The issue of inflaton fluctuations in the ``hot tin
can'' picture is not clear to us.
 In theory of inflation, scalar field fluctuations are
usually considered in the time dependent metric (\ref{flat}), as it was
described in Section \ref{sec:fluc2}. Bunch-Davies vacuum corresponds
to the thermal state in the static coordinates. On the other hand, in
the static de~Sitter coordinates, discussion of the quantum field
theory is usually restricted to thermal radiation from the horizon,
usually in terms of the detector response. Thermal radiation associated
with the de~Sitter horizon  is related to any free quantum field,
scalar fields with mass $m$ and any coupling to gravity, vector fields
etc., the difference will be only in the radial dependence of the wave
functions (``gray-body factor''). 
 On the other hand, instability of
quantum fluctuations from inflation occurs only for very light
minimally coupled scalar fields. It will be interesting to understand
what is relation of the quantum field theory in  the ``hot tin can''
picture,  and the instability of fluctuations from inflation.
For this it will be essential to work with  
regularized VEV $<\delta \phi^2>$.

As we already mentioned, in the picture of rolling scalar field with
accumulating inflaton  fluctuations, which we adopt in the paper, 
holographic bound and quantum fluctuations from inflation are compatible.

A  lesson from our calculations is that in the quasi-de~Sitter geometry
with slowly rolling scalar field, the horizon area, or geometrical
entropy, are perturbed by the scalar metric fluctuations according to
the formulas (\ref{wiggles}), (\ref{wiggles2}).
 This means that the area of the horizon
for an ensemble of the different Hubble patches is not the same but is
a statistical variable by itself. For small metric perturbations (of
order of $\Phi \sim \frac{H}{M_p}$), its mean value is $\bar
A=\frac{4\pi}{H^2}$, where $H(t)$ is slowly decreasing background value,
but dispersion around the mean value is defined by
$\langle\Phi^2\rangle$, which  growth with time. 
Equivalently, 
we can talk about statistical properties of
the local Hubble values $\tilde H$  \cite{star,stoch3}.
Notably, this
picture is converging with the stochastic description of inflation in
terms of the probability distribution of the inflaton field
\cite{stoch1,stoch2,stoch3}. Probability to have the value of inflaton
field $\phi$ in quasi-static regime 
 is $P(\phi)\sim \exp{
\left(-\frac{3M_p^4}{8V(\phi)}\right)}$ and can be interpreted in terms
of entropy $S=\log P(\phi)$ \cite{entr}. 
Remarkably, this entropy is identical to
the geometrical entropy  $\frac{A}{4G}$.
During inflation and rolling of $\phi$ the Wiener process of accumulating 
fluctuations  $\delta \phi$ (or similarly perturbations of  $\Phi$)
 changes  distribution of $P(\phi)$.
 Distribution of the horizon areas of different
Hubble patches, defined by the local values of the Hubble parameters,
depends on the background slow roll regime and the regime of
accumulation of fluctuations, both of which depend on the model of inflation
\cite{stoch1,stoch2,stoch3}.
It would be interesting to understand further the
correspondence between stochastic approach to inflation and geometrical
entropy (\ref{wiggles}), (\ref{wiggles2}).

So far we discussed small metric fluctuations $\Phi$ or small local
variations of $H$. However, in the chaotic inflationary scenario for
large enough values of $\phi$, variations of $H$ due to the quantum
jumps of $\delta \phi$ can be large and lead to the self-reproducing
inflationary universe \cite{self}.  We would like to draw attention to
this regime (which is still below the Planck energy density) and to
note that ``adiabatic'' geometrical thermodynamics which  we considered
in this paper is not applicable here. Indeed,  consider the Hubble
patch where $H$ is increasing due to the quantum jumps. Increase of
$H$ is not compatible with the classical Einstein equation
(\ref{einst}). Consequently, quantum jumps are not compatible with the
horizon thermodynamics (of a single Hubble patch)
since $\delta E=TdS$ does not hold either.
Geometrical entropy of the local Hubble patch is decreasing. 
However, we have to take into account the entropy of all Hubble
patches. Self-reproduction of inflating regions looks like a chain reaction,
which is described by the branching diffusion process \cite{branch}.

\section*{Acknowledgements}

We are grateful to Gary Felder, Valeri Frolov and Alexey Starobinsky
for  discussions and suggestions. We are especially grateful to Andrei
Linde for numerous valuable discussions
and Nemanja Kaloper for valuable discussions and clarifications.
 This research was supported by
the Natural Science and Engineering Council of Canada and the Canadian
Institute for Advanced Research.

\section*{Appendix A: Rolling Cosmic Scalars and Black Holes}

Here we briefly discuss astrophysical applications of the results of
Section \ref{sec:bh}. One of the most important examples of slowly
rolling scalar field is a homogeneous  cosmological scalar field, whose
evolution is given by the field equation $\ddot \phi +3H\dot \phi
+V_{,\phi}=0$ with the Hubble friction. This type of fields appears,
for instance, in the cosmological models of quintessence. To dominate
at the present cosmological stage while avoiding gravitational
clustering, the value of $\phi$ should be of order of the Planck mass
$M_P$, and its effective mass $m^2_{\text{eff}}=V_{,\phi\phi}$ very
small, $m^2_{\text{eff}} \leq H$, where $H(t)$ in the present day
Hubble parameter.

Quintessence may interact gravitationally with black holes of different
masses, ranging from tiny primordial black holes to astrophysical black
holes of solar masses to supermassive black holes. Accretion of
quintessence onto a black hole is given by the formula (\ref{rate1}). 
Scalar field velocity $\dot \phi(t)_{\infty}$ is defined by its
equation of motion; however, the kinetic energy  $\dot \phi^2$ cannot
exceed the total energy density of the universe $M_P^2 H^2$.
Substituting this upper limit in (\ref{rate1}), we estimate the rate of
the black hole growth due to the accretion of the rolling cosmological
scalar  field
\begin{equation}\label{estimate}
\dot M \sim H^2 M^2 \sim \left(\frac{r_g}{H^{-1}}\right)^2 \ ,
\end{equation}
where $r_g=2M$ is the gravitational radius of the black hole, $H^{-1}$
is the size of the universe, and we switched to Planck units $M_P=1$.
Thus, accretion of cosmic scalars including quintessence is absolutely
negligible for all types of astrophysical black holes.
For primordial black holes at the moment of formation, when
the ratio $ \left(\frac{r_g}{H^{-1}}\right)$ is of order of unity,
the study  of Section \ref{sec:bh} of rolling scalar field accretion 
is not applicable directly, as the black hole spacetime would be significantly
different from Schwarzschild in this case. Even if one believes the
equation (\ref{rate1}) in this regime, it would seem unlikely that
the accretion of quintessence can cause explosive growth of primordial
black holes (contrary to some claims in the literature). For this to
occur, the seed primordial black hole would have to be much bigger than
the cosmological horizon if the quintessence is subdominant.

It is interesting to take into account the back reaction of an
accretion on the evolution of the cosmic scalar field itself. From the
energy balance we obtain a correction to the Hubble friction term due
to the accretion of rolling scalar onto black holes
\begin{equation}\label{balance}
  \ddot{\phi} + (3H+ n_{\text{BH}}r_g^2) \dot{\phi} + V'(\phi) = 0 \ ,
\end{equation}
where $n_{\text{BH}}$ is the spatial density of black holes. This
correction is proportional to the filling factor of black holes, i.e.
the fraction of volume occupied by black holes. The last is tiny so
that the effect of the interaction of a rolling scalar field with black
holes is negligible.

Rolling scalar field interacting with a black hole also emerges in the
context of the Brans-Dicke or more general scalar-tensor theories
\cite{bc,jacob99}.

\section*{Appendix B: Mass Formula in Spherical Symmetry}

A general spherical spacetime is described by the metric
\begin{equation}\label{eq:ap:metric}
  ds^2 = g_{ab}\, dx^a dx^b + \rho^2 d\Omega^2,
\end{equation}
where $ g_{ab}$ is the metric on a two-manifold with coordinates
$(t, r)$, and $\rho$ is the
physical radius of spherical slices. The field dynamics is given by the
Einstein-Hilbert action, which in spherical symmetry can be
dimensionally reduced to yield
\begin{equation}
  S=\int \sqrt{-\gamma}\, d^2x\,
    \left\{ {\textstyle \frac{1}{16\pi G}} \left[\rho^2 R[\gamma] +
 2 (\nabla\rho)^2 + 2\right] + \rho^2 {\cal L}_{\text{matter}} \right\}.
\end{equation}
The Einstein equations for spherically symmetric spacetime
(\ref{eq:ap:metric}) follow from the above action by varying it with
respect to the two-metric $g_{ab}$ and the two-scalar $\rho$. In
particular, the $\{ab\}$ component of the Einstein equations is
\begin{equation}
  2\rho^{-1}[g_{ab} \Box\rho - \rho_{;ab}] + \rho^{-2} g_{ab}
 [(\nabla\rho)^2 - 1] = 8\pi G\, T_{ab}.
\end{equation}
By subtracting the (two-dimensional) trace, this equation can be
written in a more convenient form
\begin{equation}\label{eq:ap:ee}
  \rho_{;ab} - \kappa g_{ab} = -4\pi G\, \rho (T_{ab} - T g_{ab}),
\end{equation}
where we have defined
\begin{equation}\label{temp}
  \kappa = \frac{1}{2\rho} [1 - (\nabla\rho)^2] \ .
\end{equation}
In the case of Schwarzschild black hole, $\kappa$ coincides with the
surface gravity at the horizon. We will continue to use the same
notation since $\kappa$ defined in (\ref{temp}) will enter the
thermodynamics relation in the more general case. 

In spherical symmetry, it is possible to define a local mass function
$M(x^a)$ by
\begin{equation}
  1 - \frac{2GM}{\rho} = (\nabla\rho)^2 \equiv f
\end{equation}
The change in mass is related to the flux of matter given by the
stress-energy tensor $T_{ab}$. To see this, let us take the derivative
of $M=\frac{\rho}{2}\, (1-f)$
\begin{equation}
  GM_{;a} = -\frac{\rho}{2} (f_{;a} - 2\kappa\rho_{;a})
          = - \rho (\rho_{;ab} - \kappa g_{ab}) \rho^{;b}
\end{equation}
and use the Einstein equation (\ref{eq:ap:ee}) we derived above to obtain
\begin{equation}
  M_{;a} = 4\pi\rho^2 (T_{ab} - T g_{ab}) \rho^{;b}.
\end{equation}
This is the differential mass formula in a spherically symmetric
spacetime \cite{pi}.


\begin{thebibliography}{999}


\bibitem{bd}
T.~S.~Bunch and P.~C.~Davies,
{\it Quantum field theory in de Sitter space: Renormalization by point splitting},
Proc.\ Roy.\ Soc.\ Lond.\ A {\bf 360}, 117 (1978).

\bibitem{fluc}
A.~Vilenkin and L.~H.~Ford,
{\it Gravitational effects upon cosmological phase transitions},
Phys.\ Rev.\ D {\bf 26}, 1231 (1982);

\bibitem{linde}
A.~D.~Linde,
{\it Scalar field fluctuations in expanding universe and the new inflationary universe scenario},
Phys.\ Lett.\ B {\bf 116}, 335 (1982);

\bibitem{star}
A.~A.~Starobinsky,
{\it Dynamics of phase transition in the new inflationary universe scenario and generation of perturbations},
Phys.\ Lett.\ B {\bf 117}, 175 (1982).

\bibitem{metr}
V.~F.~Mukhanov and G.~V.~Chibisov,
{\it Quantum fluctuation and 'nonsingular' universe},
JETP Lett.\  {\bf 33}, 532 (1981);

\bibitem{hawking}
S.~W.~Hawking,
{\it The development of irregularities in a single bubble inflationary universe},
Phys.\ Lett.\ B {\bf 115}, 295 (1982).

\bibitem{guth}
A.~H.~Guth and S.~Y.~Pi,
{\it Fluctuations in the new inflationary universe},
Phys.\ Rev.\ Lett.\  {\bf 49}, 1110 (1982).

\bibitem{bardeen}
J.~M.~Bardeen, P.~J.~Steinhardt and M.~S.~Turner,
{\it Spontaneous creation of almost scale-free density perturbations in an inflationary universe},
Phys.\ Rev.\ D {\bf 28}, 679 (1983).

\bibitem{stoch1}
A.~Vilenkin,
{\it The birth of inflationary universes},
Phys.\ Rev.\ D {\bf 27}, 2848 (1983).

\bibitem{stoch2}
A.~A.~Starobinsky,
{\it Stochastic de Sitter (inflationary) stage in the early universe},
in {\it Current trends in field theory, quantum gravity and strings},
Eds.\ H.~de Vega and N.~Sanches,
Lect.\ Notes in Phys.\ Springer-Verlag {\bf 246} 107 (1986).

\bibitem{self}
A.~D.~Linde,
{\it Eternally existing selfreproducing chaotic inflationary universe},
Phys.\ Lett.\ B {\bf 175}, 395 (1986).

\bibitem{stoch3}
A.~S.~Goncharov, A.~D.~Linde and V.~F.~Mukhanov,
{\it The global structure of the inflationary universe},
Int.\ J.\ Mod.\ Phys.\ A {\bf 2}, 561 (1987).

\bibitem{GH}
G.~W.~Gibbons and S.~W.~Hawking,
{\it Cosmological event horizons, thermodynamics, and particle creation},
Phys.\ Rev.\ D {\bf 15}, 2738 (1977).

\bibitem{BMS}
R.~Bousso, A.~Maloney and A.~Strominger,
{\it Conformal vacua and entropy in de Sitter space},
Phys.\ Rev.\ D {\bf 65}, 104039 (2002)
[{\tt hep-th/0112218}].

\bibitem{kleban}
N.~Kaloper, M.~Kleban, A.~Lawrence, S.~Shenker and L.~Susskind,
{\it Initial conditions for inflation},
JHEP {\bf 0211}, 037 (2002)
[{\tt hep-th/0209231}].

\bibitem{kaloper} 
A.~Albrecht, N.~Kaloper and Y.~S.~Song,
{\it Holographic limitations of the effective field theory of inflation},
{\tt hep-th/0211221}.

\bibitem{jacob}
T.~Jacobson,
{\it Thermodynamics of space-time: The Einstein equation of state},
Phys.\ Rev.\ Lett.\  {\bf 75}, 1260 (1995)
[{\tt gr-qc/9504004}].

\bibitem{pi}
E.~Poisson and W.~Israel,
{\it Internal structure of black holes},
Phys.\ Rev.\ D {\bf 41}, 1796 (1990).

\bibitem{jacob99}
T.~Jacobson,
{\it Primordial black hole evolution in tensor scalar cosmology},
Phys.\ Rev.\ Lett.\  {\bf 83}, 2699 (1999)
[{\tt astro-ph/9905303}].

\bibitem{bc}
J.~D.~Barrow and B.~J.~Carr,
{\it Formation and evaporation of primordial black holes in scalar - tensor gravity theories},
Phys.\ Rev.\ D {\bf 54}, 3920 (1996).

\bibitem{mukhanov}
V.~F.~Mukhanov,
{\it Gravitational instability of the universe filled with a scalar field},
JETP Lett.\  {\bf 41}, 493 (1985).

\bibitem{KLS85}
L.~A.~Kofman, A.~D.~Linde and A.~A.~Starobinsky,
{\it Inflationary universe generated by the combined action of a scalar field and gravitational vacuum polarization},
Phys.\ Lett.\ B {\bf 157}, 361 (1985).

\bibitem{bpm}
R.~H.~Brandenberger, T.~Prokopec and V.~Mukhanov,
{\it The entropy of the gravitational field},
Phys.\ Rev.\ D {\bf 48}, 2443 (1993)
[{\tt gr-qc/9208009}].

\bibitem{Kiefer}
C.~Kiefer, D.~Polarski and A.~A.~Starobinsky,
{\it Entropy of gravitons produced in the early universe},
Phys.\ Rev.\ D {\bf 62}, 043518 (2000)
[{\tt gr-qc/9910065}].

\bibitem{entr}
A.~D.~Linde,
{\it Quantum creation of an open inflationary universe},
Phys.\ Rev.\ D {\bf 58}, 083514 (1998)
[{\tt gr-qc/9802038}].

\bibitem{branch}
A.~Linde, D.~Linde and A.~Mezhlumian,
{\it From the Big Bang theory to the theory of a stationary universe,}
Phys.\ Rev.\ D {\bf 49}, 1783 (1994)
[arXiv:gr-qc/9306035].



\end{thebibliography}
\end{document}